\documentclass[12pt,a4paper]{article}

\usepackage[T1]{fontenc}
\usepackage[iso-8859-7,latin9]{inputenc}



\usepackage[british,greek]{babel}
\begin{document}
\selectlanguage{british}
\selectlanguage{british}%
\inputencoding{iso-8859-7}\selectlanguage{british}

\selectlanguage{british}%

\title{\inputencoding{latin9}Neutrino Shortcuts in Spacetime}

\author{\inputencoding{latin9}A. Nicolaidis\medskip \\{\normalsize Theoretical
Physics Department}\\{\normalsize University of Thessaloniki}\\{\normalsize 54124
Thessaloniki, Greece}\\{\normalsize nicolaid@auth.gr}}
\maketitle
\begin{abstract}
\inputencoding{latin9}%
Theories with large extra dimensions may be tested using sterile neutrinos
living in the bulk. A bulk neutrino can mix with a flavor neutrino
localized in the brane leading to unconventional patterns of neutrino
oscillations. A resonance phenomenon, strong mixing between the flavor
and the sterile neutrino, allows to determine the radius of the large
extra dimension. If our brane is curved, then the sterile neutrino
can take a shortcut through the bulk, leading to an apparent superluminal
neutrino speed. The amount of \textquotedblleft{}superluminality\textquotedblright{}
is directly connected to parameters determining the shape of the brane.
On the experimental side, we suggest that a long baseline neutrino
beam from CERN to NESTOR neutrino telescope will help to clarify these
important issues.
\end{abstract}
\inputencoding{latin9}\qquad{}The standard model of strong and electroweak
interactions has been extremely successful. Still we know that it
cannot be regarded as the final theory. Any attempt to include gravity
leads to a unified theory where two disparate scales coexists: the
electroweak scale $\left(M_{W}\sim1\, TeV\right)$ and the Planck
scale $\left(M_{Pl}\sim10^{19}\, GeV\right)$, with an {}``energy
desert'' extended between them. Quantum corrections tend to mix the
scales and an incredible amount of fine-tuning is required (the hierarchy
problem).

A novel approach has been suggested by replacing the {}``energy desert''
with an extra dimensional space {[}1-4{]}. Our four-dimensional world
is embedded in a higher dimensional space with $D$ dimensions $\left(D=4+n\right)$.
While the standard model fields are constrained to live on the $4$-dimensional
brane, gravitons and standard model singlets, like a sterile neutrino,
can freely propagate in the higher-dimensional space (bulk). The fundamental
scale $M_{f}$ of gravity in $D$-dimensions is related to the observed
$4$-dimensional Planck scale $M_{Pl}$ by
\begin{equation}
M_{Pl}^{2}=M_{f}^{2+n}V_{n}\label{eq:1}
\end{equation}
 where $V_{n}$ is the volume of the extra space. For a torus configuration
\begin{equation}
V_{n}=\left(2\pi\right)^{n}R_{1}R_{2}\cdots R_{n}\label{eq:2}
\end{equation}
with $R_{i}$ $\left(i=1,2,\cdots,n\right)$ the radii of extra dimensions.
Then for a sufficiently large volume $V_{n}$ the fundamental scale
of gravity $M_{f}$ can be as low as $M_{W}$. In this radical way
the hierarchy problem ceases to exist as such. A phenomenology of
$TeV$ gravity has been undertaken already \cite{key-5,key-6}.

Neutrinos, from the time of their inception, some 80 years ago, remain
enigmatic and elucive. Their masses, their couplings to other particles,
the number of their flavors or types, the transformations among themselves,
remain relatively unknown. The diverse range of the neutrino parameters
allow them to play an important role from the universe evolution,
galaxy dynamics, energy generation within stars, to subatomic physics
and fundamental symmetries shaping the interactions of the elementary
particles. A sterile neutrino propagating in the bulk, may provide
informations on the extra dimensions and on properties of the brane
we are living in. It is the purpose of this letter to examine and
analyze this type of information.

A sterile neutrino propagating in the compact extra dimensions will
appear at the $4$-dimensional brane as a Kaluza - Klein (KK) tower
of states, i.e. an infinite number of $4$-dimensional spinors.The
Yukawa coupling of the standard left-handed lepton doublet, the Higgs
scalar and the right-handed bulk neutrino will provide a mixing between
the left-handed neutrino of the standard model and the KK modes. Compared
to the usual oscillations, novel features appear since now we have
a coupled system of infinite degrees of freedom. An extensive literature
on sterile neutrinos living in extra dimensions, highlights the important
aspects of an underlying rich dynamics {[}7-19{]}.

For simplicity we consider a single flavor neutrino, the muon neutrino,
and a single large extra dimenion of radius $R$. We assume all other
dimensions are smaller and do not affect our analysis. A strong mixing
between the muon neutrino and the sterile KK states is observed, whenever
a resonance condition is satisfied, an equality between the effective
mass of $\nu_{\mu}$ within matter and the mass parameters of the
KK states \cite{key-19}
\begin{equation}
\frac{G_{F}\,\varrho\, E_{\nu}}{\sqrt{2}M_{N}}=\frac{n^{2}}{R^{2}}\label{eq:3}
\end{equation}
Here $M_{N}$ is the nucleon mass, $E_{\nu}$ is the neutrino energy,
$\varrho$ is the density of the medium traversed and $n$ is referring
to the $n$th KK state (n=1, 2, 3, $\ldots$). The most significant
transition occurs for $n=1$ and the relationship (\ref{eq:3}) takes
the form

\begin{equation}
\left(\frac{\varrho}{10\, gr/cm^{3}}\right)\left(\frac{E_{\nu}}{100\, GeV}\right)\left(\frac{R}{1\,\mu m}\right)^{2}\simeq1\label{eq:4}
\end{equation}
The above condition guarantees that the transition of the muon neutrino
to the sterile neutrino and inversely, occurs with a sizeable probability.
If such a resonance phenomenon is observed, we may extract the radius
of the extra dimension. For example, for CERN neutrinos with $E_{\nu}=30\, GeV$
and an average density $\varrho$ of $5\, gr/cm^{3}$, relation (\ref{eq:4})
implies a radius $R$ of the extra dimension close to few $\mu m$.

Measurements of the neutrino time travel have been presented by MINOS
and OPERA experiments \cite{key-20,key-21}. There the possibility
of neutrinos {}``running faster than light'' has been entertained.
A number of theoretical proposals have been advanced involving newphysics,
or violations of the Lorentz symmetry {[}22-32{]}. In our work we
address the issue of the {}``superluminal'' neutrinos within the
theories of large extra dimensions, suggesting that the sterile neutrino
takes a {}``shortcut'' through the bulk. Our brane we are living
in, rather than a flat brane, may be a curved brane. On general grounds
in a brane containing matter and energy, self-gravity will induce
a curvature to the brane, so that the brane becomes concave towards
the bulk in the null direction. Then we can find geodesics in the
bulk propagating signals faster compared to the geodesics in the brane
\cite{key-33}. Standard-model neutrinos $\nu_{e},\,\nu_{\mu},\,\nu_{\tau}$
are stuck in the brane. A sterile neutrino though may leave a point
on the brane and emerge to another point on the brane, by taking the
{}``shortcut'' through the bulk and arriving earlier compared to
a flavor neutrino which follows a geodesic on the brane. A two dimensional
toy model may exhibit the expected behavior \cite{key-34,key-35,key-36}.
In a Minkowski metric 
\begin{equation}
ds^{2}=dt^{2}-dx_{1}^{2}-dx_{2}^{2}\label{eq:5}
\end{equation}
 the curved brane is represented by 
\begin{equation}
x_{2}=Asin\, kx_{1}\label{eq:6}
\end{equation}
while the bulk geodesic is given by $x_{2}=0$. A sterile neutrino
propagating through the bulk will appear as having a superluminal
speed. The time difference between the two geodesics is given by 
\begin{equation}
\frac{t_{b}-t_{s}}{t_{b}}\simeq\left(\frac{Ak}{2}\right)^{2}\label{eq:7}
\end{equation}
where $t_{b}$ ($t_{s}$) is the travel time of the flavor (sterile)
neutrino. Denoting by $v$ the effective speed of the sterile neutrino
and $c$ the speed of light, eqn. (\ref{eq:7}) takes the form
\begin{equation}
\frac{v-c}{c}=\left(\frac{Ak}{2}\right)^{2}.\label{eq:8}
\end{equation}
Thus an experimental measurement of a time difference will provide
information on the geometry of our brane, notably the brane shape
parameter $Ak$.

Within the theories of large extra dimensions, neutrinos appear as
the ideal mediators to convey information about the geometry of the
bulk and the shape of the brane. We considered the oscillation between
a flavor neutrino living in the brane and a sterile neutrino circulating
in the bulk. Our study indicates that this type of neutrino oscillations
may reveal features of the overall geometry. Further experiments are
needed to elucidate these important aspects. We suggest that a neutrino
beam from CERN directed to the NESTOR neutrino telescope, of the coast
of Pylos, will help in this direction \cite{key-37}. The large CERN-NESTOR
distance (1676 km), the different neutrino detection techniques, offer
additional leverage to study a crucial issue.

\part*{Acknowledgments}

This work was supported by the Templeton Foundation through the project
{}``Quanto-Metric''.

\end{document}